\documentclass[aps,prl,longbibliography,superscriptaddress,reprint]{revtex4-2}

\usepackage{graphicx}
\usepackage{amsmath}
\usepackage{physics}
\usepackage{braket}
\usepackage{booktabs}
\usepackage[english]{babel}
\makeatletter
\@namedef{l@en}{\l@english}
\makeatother
\usepackage[colorlinks,urlcolor=blue,citecolor=blue,linkcolor=blue]{hyperref}
\usepackage[dvipsnames]{xcolor}
\usepackage{newtxmath}

\begin{document}

\title{Quantum neural network equipped with backpropagation on a qudit processor}

\author{Yibo Yuan}\thanks{These authors contributed equally.}

\affiliation{Laboratory of Spin Magnetic Resonance, School of Physical Sciences, Anhui Provincial Key Laboratory of Scientific Instrument Development and Application, University of Science and Technology of China, Hefei 230026, China}
\affiliation{Hefei National Laboratory, University of Science and Technology of China, Hefei 230088, China}

\author{Zhuoyue Xu}\thanks{These authors contributed equally.}

\affiliation{Laboratory of Spin Magnetic Resonance, School of Physical Sciences, Anhui Provincial Key Laboratory of Scientific Instrument Development and Application, University of Science and Technology of China, Hefei 230026, China}

\author{Zhenyue Du}

\affiliation{Laboratory of Spin Magnetic Resonance, School of Physical Sciences, Anhui Provincial Key Laboratory of Scientific Instrument Development and Application, University of Science and Technology of China, Hefei 230026, China}

\author{Xingyu Zhao}

\affiliation{Laboratory of Spin Magnetic Resonance, School of Physical Sciences, Anhui Provincial Key Laboratory of Scientific Instrument Development and Application, University of Science and Technology of China, Hefei 230026, China}
\affiliation{Hefei National Laboratory, University of Science and Technology of China, Hefei 230088, China}

\author{Xu Cheng}

\affiliation{Laboratory of Spin Magnetic Resonance, School of Physical Sciences, Anhui Provincial Key Laboratory of Scientific Instrument Development and Application, University of Science and Technology of China, Hefei 230026, China}
\affiliation{Hefei National Laboratory, University of Science and Technology of China, Hefei 230088, China}

\author{Yue Li}

\affiliation{Laboratory of Spin Magnetic Resonance, School of Physical Sciences, Anhui Provincial Key Laboratory of Scientific Instrument Development and Application, University of Science and Technology of China, Hefei 230026, China}

\author{Waner Hou}

\affiliation{Laboratory of Spin Magnetic Resonance, School of Physical Sciences, Anhui Provincial Key Laboratory of Scientific Instrument Development and Application, University of Science and Technology of China, Hefei 230026, China}

\author{Yuqi Zhou}

\affiliation{Laboratory of Spin Magnetic Resonance, School of Physical Sciences, Anhui Provincial Key Laboratory of Scientific Instrument Development and Application, University of Science and Technology of China, Hefei 230026, China}

\author{Zhaokai Li}\email{zkli@ustc.edu.cn}

\affiliation{Laboratory of Spin Magnetic Resonance, School of Physical Sciences, Anhui Provincial Key Laboratory of Scientific Instrument Development and Application, University of Science and Technology of China, Hefei 230026, China}
\affiliation{Hefei National Laboratory, University of Science and Technology of China, Hefei 230088, China}
\affiliation{Hefei National Research Center for Physical Sciences at the Microscale, University of Science and Technology of China, Hefei 230026, China}

\author{Yiheng Lin}\email{yiheng@ustc.edu.cn}
\affiliation{Laboratory of Spin Magnetic Resonance, School of Physical Sciences, Anhui Provincial Key Laboratory of Scientific Instrument Development and Application, University of Science and Technology of China, Hefei 230026, China}
\affiliation{Hefei National Laboratory, University of Science and Technology of China, Hefei 230088, China}
\affiliation{Hefei National Research Center for Physical Sciences at the Microscale, University of Science and Technology of China, Hefei 230026, China}

\begin{abstract}
Quantum neural networks (QNNs), one of the fundamental algorithms in quantum machine learning, have been widely used in classification and identification tasks. However, the capabilities of QNNs are constrained by their size, which is determined by the dimension of the Hilbert space of the underlying quantum processor. Multi-level quantum digits (qudits) offer access to a higher-dimensional Hilbert space compared to two-level qubits, enabling the construction of more expressive QNNs. 
In this work, we report an experimental demonstration of qudit-based QNN using a trapped $\rm ^{40}Ca^+$ ion. 
We train the QNN using a hybrid quantum-classical implementation of backpropagation and achieve an experimental classification accuracy of $95.7\%$ on a test image set.
This demonstration highlights the potential of qudit-based processors to QNN architectures and provides a framework for implementing qudit-based QNNs across various quantum devices.
\end{abstract}


\date{\today}

\maketitle

\section{INTRODUCTION}\label{sec1}

Quantum machine learning has emerged as one of the most exciting research areas in recent years, combining quantum mechanics with machine learning techniques to explore potential quantum advantages in solving both classical and quantum learning tasks~\cite{wiebe_quantum_2012,lloyd_quantum_2014,rebentrost_quantum_2014,lloyd_quantum_2016,biamonte_quantum_2017,schuld_quantum_2019,gordon_covariance_2022,suzuki_quantum_2024, rudolph_generation_2022, dutta_single-qubit_2022}. 
Among the various algorithms in this field, quantum neural networks (QNNs) stand out for their versatility in addressing diverse problems, often implemented through parameterized quantum circuits (PQCs) ~\cite{schuld_effect_2021, beer_training_2020, pan_deep_2023, johri_nearest_2021}. QNNs have been developed for a wide range of applications and exhibit unique quantum advantages under certain conditions, including quantum data learning tasks and quantum error correction~\cite{lloyd_quantum_2018,hu_quantum_2019,cong_quantum_2019,landman_quantum_2022, liuRigorousRobustQuantum2021, abbasPowerQuantumNeural2021}.
Despite these achievements, the performance and application of QNNs are currently constrained by the limited dimension of the quantum systems, particularly with noisy intermediate-scale quantum (NISQ) devices~\cite{beer_training_2020,pan_deep_2023}.

To alleviate such limits, strategies have been explored to expand the network capacity. One approach involves leveraging models that optimize quantum resource usage, such as re-uploading models to construct wider networks~\cite{schuld_effect_2021}. Another promising strategy employs multi-level quantum systems, or qudits, as processor units for QNNs. Unlike traditional two-level quantum bits (qubits), qudits offer access to a higher-dimensional Hilbert space, represented by a capacity of $d^n$, where $n$ is the number of qudits and $d$ is the number of available levels in one qudit~\cite{wach_data_2023, useche_quantum_2021, mandilara_classification_2024}. Therefore, using a qudit instead of a qubit can potentially enhance the quantum capabilities of neural networks. Moreover, qudits naturally exist on various experimental platforms with multi-level structures, as recently demonstrated in trapped-ions~\cite{low_practical_2020,ringbauer_universal_2022,low_control_2025,zalivako_towards_2025}, photonic systems~\cite{chi_programmable_2022,du_qudit-based_2024,kues_-chip_2017}, and superconducting quantum systems~\cite{kristen_amplitude_2020,tan_experimental_2021,zheng_optimal_2022,kiktenko_single_2015}.

\begin{figure*}[ht]
\includegraphics[width=14cm]{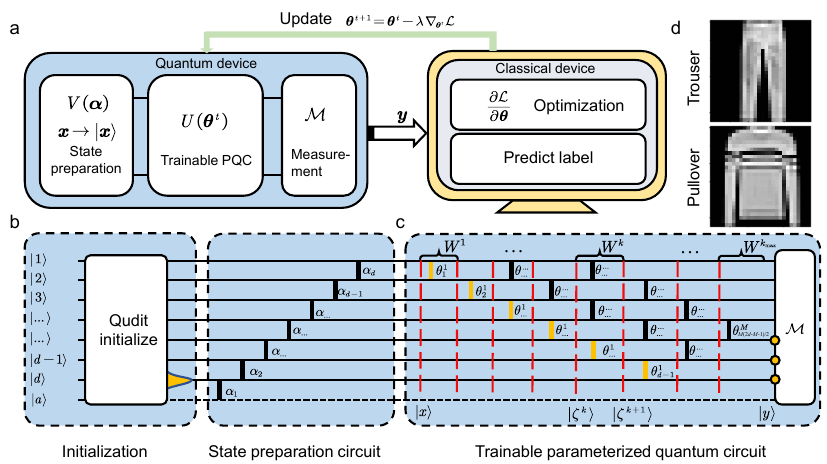}
\caption{(a) Schematic diagram of the qudit QNN. (b) The qudit initialization and state preparation circuits. The Qudit processor is initialized into level $\ket{d}$ at the beginning. The state preparation circuit maps the input data vector $\boldsymbol{x}$ into the data state $\ket{x}$. The vertical line represents a transition $T(\alpha)$ described in the main text.
(c) The trainable circuit. This quantum circuit is composed of a series of stairs. The orange lines form the first transition stair, and the orange dots indicate the levels whose amplitudes are used to predict the label of the input data. The circuit is divided into several sublayers by red dashed lines. Each sublayer $W^k$ corresponds to an inner data $\boldsymbol{\zeta}^k$ and an inner error $\boldsymbol{\delta}^k$. (d) Two samples in the Fashion-MNIST dataset used in the experiment.
}
	\label{fig: algorithm diagram and circuit}
\end{figure*} 

In this work, we focus on the theoretical design and experimental realization of neural networks on a trapped-ion qudit processor. By encoding quantum information in the energy levels and utilizing transitions between qudit levels, we adapt a PQC into a qudit-based framework. 
Trained using a hybrid quantum-classical implementation of the backpropagation~\cite{landman_quantum_2022,kerenidis2022classicalquantumalgorithmsorthogonal,beer_training_2020,pan_deep_2023}, our implementation with a trapped $\mathrm{^{40}Ca^+}$ ion achieves a test accuracy of $95.7\%$ in the experiment.
This result demonstrates the feasibility of constructing QNNs on qudit-based hardware. Moreover, our approach is readily extendable to multi-qudit architectures, enabling scalable implementations for practical applications.

\section{ALGORITHM}\label{sec2}

We demonstrate a QNN implemented on a qudit processor through a multi-class ($M$-class) classification task with $d$-dimensional input data, aiming to predict the category label of the inputs. The network on a qudit consists of a data state preparation circuit, a trainable PQC, and supporting components, as schematically shown in Fig.~\ref{fig: algorithm diagram and circuit}a, 
in which the trainable parameters are optimized using the hybrid quantum-classical backpropagation.
The basic operations in the qudit network are parameterized transitions between pairs of adjacent levels $\{\ket{i},\ket{i+1}\}$, denoted as $T_{i,i+1}(\theta)=\begin{bmatrix} 
                     \cos{\frac{\theta}{2}}& \sin{\frac{\theta}{2}}&\\
                     -\sin{\frac{\theta}{2}}& \cos{\frac{\theta}{2}}&\\
\end{bmatrix}_{i,i+1}$.
To encode the data into quantum states, we adopt the amplitude encoding method commonly used in quantum algorithms~\cite{schuld_quantum_2019}. The input data vector is labeled as $ \boldsymbol{x} = (x_1, \dots, x_d)^T \in \mathbb{R}^d$. The signs of the vector elements can be determined by referring to an ancillary level with positive amplitude $x_a\in(0, 1)$, thus we encode these amplitudes into the state
\begin{equation}
\ket{{x}} =V(\boldsymbol{\alpha})\ket{d} = \sum_{i=1}^d x_i\ket{i} + x_{a}\ket{a},
    \label{Eq:generate_psi_0}
\end{equation}
where $\ket{i}$ and $\ket{a}$ are the respective qudit and ancillary levels with the input amplitudes normalized such that $|\boldsymbol{x}|^2 = 1 - x_a^2$. The qudit is first initialized to $\ket{d}$, then the encoding operation $V(\boldsymbol{\alpha})=\prod_{j=2}^{j=d}T_{d-j+1,d-j+2}(\alpha_j)T_{d,a}(\alpha_1)$ is applied via applying a qudit circuit, as shown in Fig.~\ref{fig: algorithm diagram and circuit}b. The parameters $\boldsymbol{\alpha}=(\alpha_1,...,\alpha_d)^T$ are determined from the data vector $\boldsymbol{x}$ (details can be found in the Supplementary Data). 

For the $M$ classes classification task, the amplitudes on levels $\{\ket{d-M+1},...,\ket{d-1},\ket{d}\}$ are used to predict the label of input data. 
To connect those $M$ levels with the input levels, a trainable PQC with $M$ transition stairs is built, and there are $d-l$ transitions in the $l$-th transition stair~\cite{roca-jerat_qudit_2024, kerenidis2022classicalquantumalgorithmsorthogonal, landman_quantum_2022}, as illustrated in Fig.~\ref{fig: algorithm diagram and circuit}c. Thus, the trainable circuit contains a total of ${M(2d-M-1)/2}$ independent rotation parameters, denoted by $\boldsymbol{\theta}=(\theta_1^1, ..., \theta_j^l, ..., \theta_{d-M}^M) \in \mathbb{R}^{M(2d-M-1)/2}$, in which $\theta_j^l$ represents the $j$-th rotation parameter in the $l$-th transition stair.
For example, the first transition stair is given by $\prod_{j=1}^{d-1}T_{j,j+1}(\theta_j^1)$, as highlighted in Fig.~\ref{fig: algorithm diagram and circuit}c. 
The unitary matrix of the trainable circuit is given by $U(\boldsymbol{\theta})$ as 
\begin{equation}
U(\boldsymbol{\theta})=\prod_{l=1}^{M} \prod_{j=1}^{d-l}T_{j,j+1}(\theta_j^l).
    \label{Eq:trainable_pqc}
\end{equation}

At the end of the circuit, the output state is $\ket{{y}} = U(\boldsymbol{\theta})\ket{x} = \sum_{i=1}^d y_i\ket{i} + x_{a}\ket{a}$.
Here $\boldsymbol{y}=(y_1,...,y_d)^T$ is the output vector of the QNN, which can be extracted by measuring the output state $\ket{{y}}$.

For classification tasks, neural networks usually employ the softmax function to convert output vectors into confidence probabilities of labels. For the sample $\boldsymbol{x}$, after applying the quantum circuit, the output vector $\boldsymbol{y}$ can be obtained. For this $M$ classes classification task, the set of labels is $[M]=\{1,2,..., M\}$. We use components $\{y_{d-M+1},...,y_d\}$ of $\boldsymbol{y}$ to compute the probability of each label. For label $i\in[M]$, the probability is
\begin{equation}
p_i = \frac{e^{y_{d-M+i}}}{\sum^{d}_{j=d-M+1} e^{y_j}}.
    \label{Eq:probability_i}
\end{equation}
Cross-entropy is then used as the loss function. Assume the actual label of the sample is $t\in[M]$, then the loss $\mathcal{L}$ for this sample is
\begin{equation}
\mathcal{L}=-\ln(p_t) +\gamma\sum_{i=1}^{d-M}|y_i|^2.
    \label{Eq:loss_definition}
\end{equation}
Here, $p_t$ is the probability that the class of the sample is correctly predicted by the QNN. In addition, a regular term $\sum_{i=1}^{d-M}|y_i|^2$ in the loss function is used to reduce the amplitudes on levels that are not used for label predicting, and $\gamma$ denotes the weight assigned to the regular term~\cite{huang_realization_2020}. 
During the training process, $\mathcal{L}$ is minimized to improve the prediction ability of the QNN.

The QNN is trained via gradient descent, with gradients evaluated through backpropagation. 
To perform the backpropagation, we divide the quantum circuit into $k_{max}$ sublayers in the form 
$U(\boldsymbol{\theta})=\prod_{k=1}^{k_{max}}W^k$, as shown in Fig.~\ref{fig: algorithm diagram and circuit}c.
The amplitude $x_a$ on the ancillary level $\ket{a}$ is not changed after applying the quantum circuit. Thus, we omit the ancillary level $\ket{a}$ and use $\hat{U}(\boldsymbol{\theta})=\prod_{k=1}^{k_{max}}\hat{W}^k$ to describe the computation of data vectors in the trainable quantum circuit, which can be divided into sublayers. The inner data vector $\boldsymbol{\zeta}^k$ within the circuit is measured from its corresponding data state $\ket{\zeta^k}$ and satisfies the relation $\boldsymbol{\zeta}^{k+1} = \hat{W}^k \boldsymbol{\zeta}^k$. 
For parameter $\theta_j^l$ in sublayer $W^k$, the gradient is 
\begin{equation}
\begin{aligned}
    \frac{\partial\mathcal{L}}{\partial\theta_j^l}
    &=\frac{\partial\mathcal{L}}{\partial\boldsymbol{\zeta}^{k+1}}\frac{\partial\boldsymbol{\zeta}^{k+1}}{\partial\theta_j^l}
    =\frac{\partial\mathcal{L}}{\partial{\boldsymbol{\zeta}^{k+1}}}\frac{\partial \hat{W}^k(\theta_j^l)\boldsymbol{\zeta}^{k}}{\partial\theta_j^l}
    \\
    &=(\boldsymbol{\delta}^{k+1})^T\frac{\partial \hat{W}^k(\theta_j^l)}{\partial\theta_j^l}\boldsymbol{\zeta}^k.
\end{aligned}
    \label{Eq:chain_rule}
\end{equation}

Here we define the inner error vector $\boldsymbol{\delta}^{k+1}=(\partial\mathcal{L}/\partial\boldsymbol{\zeta}^{k+1})^T$. 
With the chain rule, the inner error vectors satisfy the backward recursive relation $\boldsymbol{\delta}^{k}=\hat{W}^{k\dagger}\boldsymbol{\delta}^{k+1}$.
The output error vector is first computed as $\boldsymbol{\Delta}=\left( \partial\mathcal{L}/\partial \boldsymbol{y}\right)^T$. The error vector at the $k_{\max}$th inner layer is then given by $\boldsymbol{\delta}^{k_{\max}} = \hat{W}^{k_{\max}\dagger} \boldsymbol{\Delta}$. Furthermore, all the inner error vectors $\{\boldsymbol{\delta}^1, ..., \boldsymbol{\delta}^{k_{max}}\}$ for the sublayers can be retrieved through the backward recursive relationship.
These inner error vectors, along with the inner data vectors, are then used to compute the gradients of the parameters on classical devices.
As a potential extension, a fully quantum backpropagation could be realized by encoding the error vectors as quantum states and propagating them backward through the quantum circuit (see Supplementary Data).

After acquiring the gradient in each sublayer, all the trainable parameters in the QNN can be updated iteratively, with
\begin{equation}
    \boldsymbol{\theta}^{h+1}=\boldsymbol{\theta}^{h}-\lambda\left({\partial\mathcal{L}/\partial \boldsymbol{\theta}^h}\right)^T.
    \label{Eq:parameters_update}
\end{equation}
Here, $\lambda$ is the learning rate set to control the step length of training, and the superscript $h$ is the index of the training epoch. Finally, the trained neural network outputs the label with the highest confidence probability as the predicted label of the input data.

\section{EXPERIMENTAL SETUP}\label{sec3}

\begin{figure}[t]
\includegraphics[width=6.5cm]{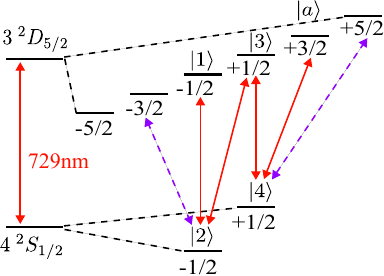}
\caption{The energy levels of the $^{40}\rm{Ca}^+$ ion. $\ket{1}$,$\ket{2}$,$\ket{3}$,$\ket{4}$ are used to encode the data vector $\boldsymbol{x}$ and $\ket{a}$ is the ancillary level used for sign readout. The two 729~nm sub-transitions to assist detection of multiple levels~\cite{nagourney_shelved_1986} are denoted by purple dashed arrows.}
\label{fig:energy_level_and_exp_pulse}
\end{figure}

Our experiment is demonstrated on a single trapped $^{40}\rm{Ca}^+$ ion in a linear Paul trap with an ambient magnetic field of 0.54~mT. We utilize internal levels to construct a qudit system with $d=4$, where $\ket{1}$, $\ket{2}$, $\ket{3}$, $\ket{4}$, $\ket{a}$ levels are the fine structure sub-levels $\ket{D_{5/2},-1/2}$, $\ket{S_{1/2},-1/2}$, $\ket{D_{5/2},+1/2}$, $\ket{S_{1/2},+1/2}$, $\ket{D_{5/2},+3/2}$, respectively, with the ground levels $S_{1/2}$ and the metastable levels $D_{5/2}$ of the $^{40}\rm{Ca}^+$ ion as depicted in Fig.~\ref{fig:energy_level_and_exp_pulse}. The transition between $S_{1/2}$ and $D_{5/2}$ can be resonantly driven by a narrow linewidth 729~nm laser, which implements the coherent manipulation $T(\cdot)$. 
These transitions can be frequency-resolved and thus can be driven by dynamically tuning the strength, frequency, and phase of the 729~nm laser with an acousto-optic modulator~\cite{roos_controlling_nodate,ringbauer_universal_2022}. In the experiment, the fidelities of the single-qudit $T(\cdot)$ operations exceed $99\%$.
At the beginning of the experiment, a series of controlled 397, 866, 854, and 729~nm laser beam pulses are applied to the ion to cool its motion and initialize it to $\ket{4}$, and then we use 729~nm laser pulses to apply unitary evolutions, including the state preparation circuit and the trainable circuit. 

We apply fluorescence detection for readout, where we typically apply 397 and 866~nm lasers for 300 $\mu$s to collect an average of 36.4 (2.7) counts when the ion is on the $S_{1/2}$ ($D_{5/2}$) manifold. 
We apply a sequential measurement~\cite{ringbauer_universal_2022,leupold_sustained_2018} to acquire all five populations on the qudit levels with one experimental sequence, which reduces the time overhead (see Supplementary Data).
The state preparation and measurement error for dark states are approximately $0.2\%$ in the experiment.
The magnitude of each amplitude is extracted from qudit populations, and its sign is determined through interference with the ancillary level $\ket{a}$. We follow the measurement process similar to the population readout to measure the signs, with the addition of $\uppi/2$ pulses interleaved within the detection process. In practice, we take 1500 shots to measure all the populations and 600 shots for each sign.

As a demonstration, we implement a QNN with four-dimensional input to classify trousers ($\mathcal{T}$) and pullovers ($\mathcal{P}$) from the Fashion-MNIST dataset~\cite{Fashion}. To match the qudit capacity, we use the Principal Component Analysis~\cite{grant_hierarchical_2018,landman_quantum_2022} method to compress the original $28\times28$ images to $4\times1$ vectors. 
The first four principal components accounted for 73.69\% of the explainable variance in the training set. After the PCA transformation had been fixed, every test image was centered using the training set mean and projected onto the same four principal components, without refitting the PCA model.

\begin{figure}[t]
\centering
\includegraphics[width=6.5cm]{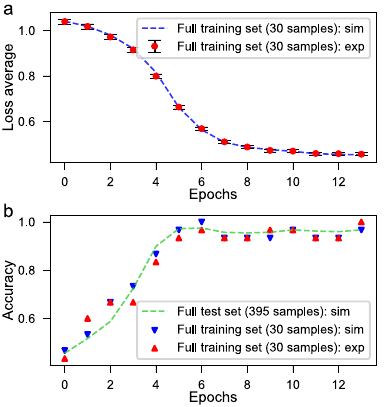}
\caption{(a) Average loss. 
The red dots denote experimental results, while the blue line represents noise-free numerical simulations performed using the experimentally trained parameters $\boldsymbol{\theta}$. The average is computed over all 30 training samples. (b) Classification accuracy. Red dots denote experimental results, while blue dots and the green dashed line represent numerical simulations using the experimentally trained parameters $\boldsymbol{\theta}$.
}
\label{fig:single_layer_result_1}
\end{figure} 

In the experiment, a training set of 30 samples (15 $\mathcal{T}$ and 15 $\mathcal{P}$) is used to train our QNN. The trainable parameters are initialized to zeros ($\boldsymbol{\theta}= \boldsymbol{0}$). The choice of initialization does not significantly affect the training dynamics or final performance (see Materials and Methods).
The output vector $\boldsymbol{y}$ and intermediate data vectors $\{\boldsymbol{\zeta}^k\}$ are obtained through measurements on the qudit processor. Based on these measured quantities, the loss $\mathcal{L}$, error vectors $\{\boldsymbol{\delta}^{k}\}$, and gradients $\{\partial\mathcal{L}/\partial\theta_j^l\}$ are computed on a classical device~\cite{kerenidis2022classicalquantumalgorithmsorthogonal,landman_quantum_2022}.
In each epoch, 15 samples are randomly selected from the training set, and their accumulated gradient is used to update the parameters. The network is trained for 13 epochs with a learning rate $\lambda=0.2$ and a regular term weight $\gamma=0.5$. The final test set accuracy is insensitive to the choice of $\gamma$ (see Supplementary Data).

\begin{figure*}[ht]
    \centering
	\includegraphics[width=14cm]{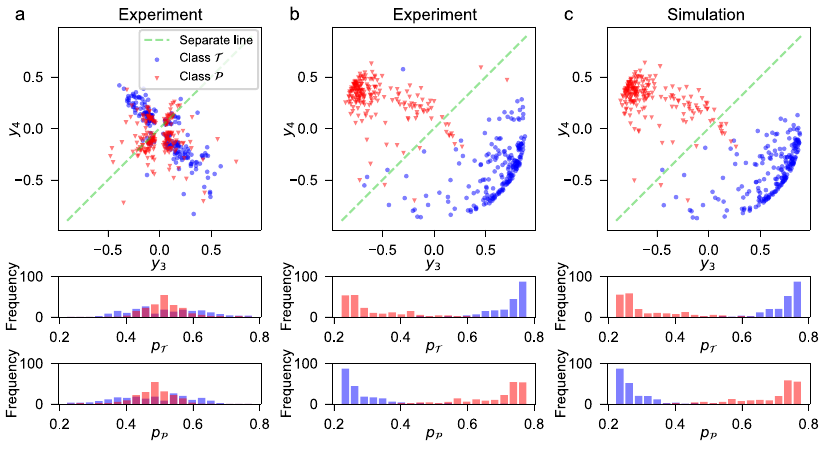}\\
	\caption{The distribution of the last two components $y_3$, $y_4$ of the output vector of the qudit neural network. These components are used to classify the correlated samples. The experimental results before training (a) and after the training process (b). (c) The numerically simulated result using the parameters $\boldsymbol{\theta}$ obtained through the experiment. The red and blue symbols represent the two classes of images in a 395-sample test set. Histograms show the one-dimensional distributions of $p_i$. After 13 epochs of training, the classification accuracy is improved from $45.6\%$ to $95.7\%$ in the experiment.}
	\label{fig:single_layer_result_3}
\end{figure*}

\section{RESULTS}\label{sec4}

The experimental results during the training process are shown in Fig.~\ref{fig:single_layer_result_1}, including average training loss and classification accuracy at each training epoch. For comparison, the figure also presents the noise-free numerical simulations of the training set loss and accuracy using the parameters obtained experimentally at each epoch.
Because the training set contains only 30 samples, a change in the classification of a single sample corresponds to observed discrete fluctuations in training accuracy.
Before the experimental evaluation, we randomly selected a test set from the Fashion-MNIST dataset. After excluding samples lacking measurement data due to ion loss in the experiment, the resulting test set consisted of 395 samples, comprising 197 trousers ($\mathcal{T}$) and 198 pullovers ($\mathcal{P}$).
To limit the experimental overhead, the full 395-sample test set was measured experimentally only before training and after the final training epoch. At the intermediate epochs, we performed noise-free numerical simulations to evaluate the test set accuracy using the experimentally trained parameters.
The accuracy on the test set is close to 1 after the fifth epoch and reaches $96.7\%$ after the $13$-th epoch, as shown in Fig.~\ref{fig:single_layer_result_1}b. The numerical test set accuracy indicates that the QNN can be efficiently trained and exhibits strong generalization on unknown data.

Visualizing the neural network output provides insight into its classification behavior.
In the experiment, amplitudes $(y_3,y_4)$ on levels $(\ket{3},\ket{4})$ are used to evaluate the confidence probabilities of the labels. 
Fig.~\ref{fig:single_layer_result_3} shows how the values of these two components evolve during the training process, along with the corresponding distribution of confidence probabilities $(p_{\mathcal{T}},p_{\mathcal{P}})$. 
Initially, the two data classes are indistinguishable (Fig.~\ref{fig:single_layer_result_3}a). After training, a distinct classification boundary emerges, resulting in a significant separation between the two classes of images (Fig.~\ref{fig:single_layer_result_3}b).
Numerical classification using parameters $\boldsymbol{\theta}$ obtained from the experiment is shown in Fig.~\ref{fig:single_layer_result_3}c, exhibiting good agreement with the experimental results.
The residual analysis of $(y_3,y_4)$ can be found in Supplementary Data.
The difference between the experimental and numerical results is due to imperfections in the experiment, which we attribute mainly to slow and fast environmental magnetic field noise (details can be found in Supplementary Data). In addition, due to the finite size of the measurement, the absolute values of $\boldsymbol{y}$ and $\{\boldsymbol{\zeta}^1,...,\boldsymbol{\zeta}^{k_{max}}\}$ would experience a standard deviation of about $0.013$ given 1500 shots for each experiment.

\section{DISCUSSION}\label{sec5}

In this work, we introduced a qudit-based quantum neural network equipped with backpropagation. By utilizing multiple energy levels in qudits, machine learning models can achieve enhanced expressivity and learning capacity without increasing the size of the quantum system. 
In the experiment, we encode the data into several fine-structure energy levels of the $^{40}\rm{Ca}^+$ ion, and implement the QNN for an image classification task on the Fashion-MNIST dataset. The network achieved a test accuracy of $95.7\%$ after 13 training epochs, closely matching the numerical simulation ($96.7\%$). Our results demonstrate that multi-level quantum systems are well-suited for machine learning tasks with limited physical resources~\cite{low_practical_2020}, and illustrate that QNNs can be scaled with significantly reduced experimental complexity.

To further enhance model capacity, deep QNNs can be constructed by introducing nonlinearity through activation functions. These nonlinearities can be implemented via block encoding based on quantum singular value transformation~\cite{guo_nonlinear_2024, guo_quantum_2024}, or by using a hybrid quantum-classical framework where activation functions are computed on a classical device~\cite{HQCNN, LIANG2021133, liuHybridQuantumclassicalConvolutional2021}.  Additionally, recently developed qudit shadow estimation techniques offer a route to significantly reduce the training cost of backpropagation, making qudit-based QNNs more practical for near-term applications~\cite{mao_qudit_2025, abbas_quantum_2023}.
These directions highlight the potential of qudit-based QNNs, positioning them as a powerful platform for scalable quantum machine learning.

\section{MATERIALS AND METHODS\label{sec6}}

\subsection{Backpropagation\label{app1}}

Using the chain-rule relation in Eq.~(\ref{Eq:chain_rule}), we consider parameter $\theta_j^l$ of the transition $T_{i,i+1}(\theta_j^l)$ within $\hat{W}^k$. In the matrix representation, the derivative $\partial \hat{W}^k(\theta_j^l)/\partial\theta_j^l$ is nonzero only within the $2\times2$ block associated with the subspace spanned by $\{\ket{i},\ket{i+1}\}$. Thus, only the components within this subspace contribute to the gradient $\partial\mathcal{L}/\partial\theta_j^l$, which can be written as

\begin{equation}
    \begin{aligned}
    \frac{\partial \mathcal{L}}{\partial \theta_j^l} &= 
    \boldsymbol{\delta}^{k+1}[i] \frac{\partial}{\partial \theta_j^l}(\cos{\frac{\theta_j^l}{2}}\boldsymbol{\zeta}^k[i]      + \sin{\frac{\theta_j^l}{2}}\boldsymbol{\zeta}^k[i+1])\\
    &+\boldsymbol{\delta}^{k+1}[i+1] \frac{\partial}{\partial \theta_j^l}(-\sin{\frac{\theta_j^l}{2}}\boldsymbol{\zeta}^k[i] \\
    &+ \cos{\frac{\theta_j^l}{2}}\boldsymbol{\zeta}^k[i+1]),
    \label{parameter_gradient}
    \end{aligned}
\end{equation}
or
\begin{equation}
    \begin{aligned}
    \frac{\partial \mathcal{L}}{\partial \theta_j^l} &=     \frac{1}{2}\boldsymbol{\delta}^{k+1}[i](-\sin{\frac{\theta_j^l}{2}}\boldsymbol{\zeta}^k[i]      + \cos{\frac{\theta_j^l}{2}}\boldsymbol{\zeta}^k[i+1])\\
    &+\frac{1}{2}\boldsymbol{\delta}^{k+1}[i+1] (-\cos{\frac{\theta_j^l}{2}}\boldsymbol{\zeta}^k[i] \\
    &-\sin{\frac{\theta_j^l}{2}}\boldsymbol{\zeta}^k[i+1]).
    \label{parameter_gradient_2}
    \end{aligned}
\end{equation}
Here $\boldsymbol{\zeta}^k[i]$ and $\boldsymbol{\delta}^k[i]$ denote the $i^{th}$ components of the data and error vectors, respectively~\cite{landman_quantum_2022, kerenidis2022classicalquantumalgorithmsorthogonal}.

\subsection{Baseline comparisons\label{app_baseline}}

We present several baseline comparisons including a linear support vector machine, a two-qubit QNN, and a noise-free numerical simulation, as shown in Table~\ref{tab:baseline_comparisons}. 
Note that the numerical simulation of the single-qudit QNN follows an independent training trajectory and should be distinguished from the numerical simulations in Fig.~\ref{fig:single_layer_result_1} and Fig.~\ref{fig:single_layer_result_3}, which were obtained using the experimentally trained parameters.
Table~\ref{tab:baseline_comparisons} shows that the three models achieve comparable test performance on this small-scale proof-of-principle task. Further details can be found in the Supplementary Data. 

\begin{table*}[ht]
    \centering
    \caption{Baseline Comparison}
    \label{tab:baseline_comparisons}
    \begin{tabular}{l c c c}
        \toprule
        \textbf{Model} & \textbf{Training Set Accuracy} & \textbf{Test Set Accuracy} \\
        \midrule
        Linear SVM      & 100.0\%            & 97.2\%            \\
        Two-qubit QNN      & 100.0\%            & 95.4\%            \\
        Single-qudit QNN (sim)      & 96.7\%            & 96.7\%            \\
        Single-qudit QNN (exp)      & 100.0\%          & 95.7\%            \\
        \bottomrule
    \end{tabular}
\end{table*}

\subsection{The training dynamics of the parameters\label{app2}}

In the experiment, all the parameters are initialized to zero. To systematically assess whether this choice influences training dynamics or final performance, we conducted a comprehensive numerical study involving: 
(i) Sampling 1000 distinct random parameter initializations of $\boldsymbol{\theta}$.
(ii) Performing 13 training iterations for each configuration with a fixed learning rate.

\begin{figure*}[ht]
\includegraphics[width=14cm]{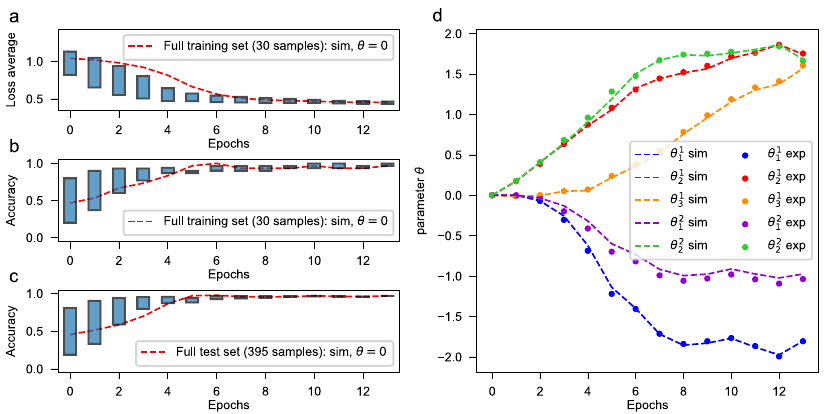}
\caption{
 Convergence of the average loss (a), training set accuracy (b), and test set accuracy (c) during the training process, based on 1,000 distinct parameter initializations. The red dashed line represents the situation where the initial parameters are set to zero. In the box plots, the bottom and top edges represent the first (Q1) and third (Q3) quartiles, respectively. (d) The training path of the parameters $\boldsymbol{\theta}$ with zero-initialized scheme. The $i$-th dot denotes  $\theta$ calculated after the $i$-th training epoch. 
	}
	\label{fig:loss_and_accuacy_random_1000_and_theta_path}
\end{figure*}

As shown in Fig.~\ref{fig:loss_and_accuacy_random_1000_and_theta_path}, the QNN consistently converges to a test accuracy exceeding $95\%$ regardless of the initialization. The average loss stabilizes around 0.5, demonstrating reliable convergence behavior. Notably, the training dynamics and performance under random initializations closely mirror those of the zero-initialized case. This suggests that zero initialization does not restrict access to favorable regions of the solution landscape in our setup. Overall, these results indicate that our QNN training is robust to the choice of parameter initialization and that using zero-initialized parameters does not adversely impact performance in this context. Based on the zero-initialized scheme, the training path of the parameters $\boldsymbol{\theta}$ is shown in Fig.~\ref{fig:loss_and_accuacy_random_1000_and_theta_path}d, where the simulation and experiment exhibit similar trends and final performance.

\section{ACKNOWLEDGMENTS\label{ACKNOWLEDGMENTS}}
This work was supported by the Quantum Science and Technology-National Science and Technology Major Project (2024ZD0302000 and 2021ZD0301603), National Natural Science Foundation of China (92565306, 92165206 and 92165108),  the Chinese Academy of Sciences (XDB1300000), and the National Key Research and Development Program of China (2025YFE0217900).

\bibliographystyle{apsrev4-2}

\clearpage
\onecolumngrid

\begin{center}
{\large\bfseries Supplemental Material for\\
“Quantum neural network equipped with backpropagation on a qudit processor”}
\end{center}

\section{\label{section0}Appendix A: Loss design and full quantum backpropagation}

Cross-entropy loss is adopted as the objective function for the classification task. 
For a given sample with the target label $t\in[M]$, the loss function is defined as
\begin{equation}
\mathcal{L}
=
-\ln(p_t)
+
\gamma\sum_{i=1}^{d-M}|y_i|^2,
\label{Eq:loss_definition}
\end{equation}
where the second term introduces a regularization penalty for the amplitudes of the unassigned output components. The parameter $\gamma$ controls the strength of this regularization term.

\begin{figure}[ht]
\centering
    \includegraphics[width=0.49\columnwidth]{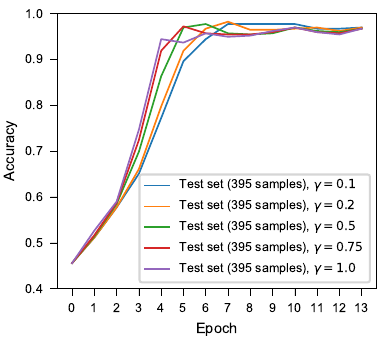}
  \caption{Classification accuracies for different values of the regularization coefficient $\gamma$ during the training process.}
  \label{fig:gamma_analyze}
\end{figure}

To examine the sensitivity of the model performance to the regularization weight $\gamma$, we performed a sensitivity analysis by varying $\gamma$ from 0.1 to 1.0, as shown in Fig.~\ref{fig:gamma_analyze}. The value $\gamma=0.5$ used in our simulations was chosen as a representative intermediate value, without any dedicated hyperparameter optimization. The results show that the final classification accuracy remains stable over a broad range of $\gamma$, indicating that the model performance is insensitive to the precise choice of the regularization strength $\gamma$.

A fully quantum implementation of the backpropagation procedure can be constructed by encoding the error vectors into quantum states. Similar to the amplitude encoding of the input data vector, the normalized error vector $\boldsymbol{\Delta}$ can be encoded as an error state.
To encode the error vector into a quantum state, the error vector 
$\boldsymbol{\Delta}$ is normalized with $x_a$ and the norm of the original vector $\|\boldsymbol{\Delta}\|$ is recorded as the normalization factor during the encoding process. 
The normalized error state is propagated backward through the inverse of the trainable quantum circuit, which corresponds to the reverse evolution of the forward propagation process. By measuring the intermediate error states generated during this backward propagation, the required error vectors $\{\boldsymbol{\delta}\}$ can be reconstructed on the quantum processor. Specifically, the measured amplitudes are multiplied by the stored normalization factor $\|\boldsymbol{\Delta}\|$ to recover the original scale of the error vectors. Consequently, the gradients of the trainable parameters can be evaluated without explicitly computing the error vectors on a classical computer. This approach provides a possible fully quantum extension of the backpropagation protocol, referred to as quantum backpropagation.

\section{\label{section1}Appendix B: Data encoding }

For a reduced Fashion-MNIST data vector $X=(X_1, X_2, X_3, X_4)$, in which $\Sigma X_i^2 = 1$, we use amplitude encoding to map it onto the $\ket{1}, \ket{2}, \ket{3}, \ket{4}, \ket{a}$ levels, where $\ket{a}$ is the ancillary level. The resulting quantum state is 
\begin{equation}
    \ket{\boldsymbol{x}}=x_1\ket{1}+x_2\ket{2}+x_3\ket{3}+x_4\ket{4}+x_a\ket{a},
    \label{Eq:data_after_e_process}
\end{equation}
where the amplitudes $x_i$ are defined as
\begin{equation}
x_i=X_i\sqrt{1-x_a^2}~\ (i=1,2,3,4),
\label{Eq:data_amp_after_e_process}
\end{equation} 
with $x_a = \sqrt{0.2}$ in the experiment.

To prepare the quantum state $\ket{\boldsymbol{x}}$ from the initial state $\ket{4}$, four 729~nm laser pulses, labeled $P_1, P_2, P_3, P_4$, need to be applied. We choose y-pulses to ensure the amplitudes are real numbers. The signs of the elements in the data vector play a crucial role in the encoding process. For example, two different sets of pulses are required for the following two data vectors:
\begin{equation}
\begin{aligned}
&X_{v1}=(+0.5,+0.5,+0.5,+0.5),\\
&X_{v2}=(-0.5,+0.5,-0.5,+0.5).
    \label{Eq:two_data_vectors}
\end{aligned} 
\end{equation} 

In Rabi oscillations between two energy levels, the quantum state changes periodically if no errors occur, with the population changing twice as fast as the phase. We define the $\pi$ time of population inversion in a Rabi oscillation between levels $\ket{m}$ and $\ket{n}$ as $T_{\pi} = T_{m,n}$. Specifically, after $2T_{\pi}$, the population returns to its initial value, while the phase (or the sign of the amplitude) requires $4T_{\pi}$ to return to its initial state. 

The initial states corresponding to the two data vectors described in Eq.~(\ref{Eq:two_data_vectors}) have the same population but different phases. Traditionally, if the quantum state always rotates around the y-axis of the Bloch Sphere with the same rotation direction, one of the vectors will take longer to prepare than the other. While this might not matter theoretically, longer preparation times introduce more experimental errors due to decoherence. Therefore, we set the quantum state to always rotate around the y-axis, with the direction of rotation determined by the sign of the data vector, either in the positive or negative direction. Ultimately, the preparation time for all pulses $P_i$ is limited to $T_{\pi}$, which greatly reduces the effects of decoherence. The two data vectors in Eq.~(\ref{Eq:two_data_vectors}) share the same preparation time, but their pulse phases differ.

We set the amplitude of the ancillary level $\ket{a}$ to always be positive, i.e., always $+x_a$. Assuming the phases of the pulses in the parameterized quantum circuit are 0, the phases of the pulses used in the initial state preparation will be set to 0 or $\pi$, depending on the signs of the data vectors. For brevity, a detailed discussion of each pulse phase is omitted here. The preparation times for the pulses $P_i$ can be calculated as
\begin{equation}
\begin{aligned}
t_1&=\arcsin(x_a)\frac{2T_{4,a}}{\pi},
\\
t_2&=\arccos(\abs{X_4})\frac{2T_{3,4}}{\pi},
\\
t_3&=\arccos[\abs{X_3} /\sin(\frac{\pi t_2}{2T_{3,4}})
]\frac{2T_{2,3}}{\pi},
\\
t_4&=\arccos[\abs{X_2} /\sin(\frac{\pi t_2}{2T_{3,4}})/\sin(\frac{\pi t_3}{2T_{2,3}})
]\frac{2T_{1,2}}{\pi}.
    \label{Eq:prepare_time}
\end{aligned} 
\end{equation} 

Alternatively, if the pulse lengths are expressed in terms of rotation angles, we have the following relations:
\begin{equation}
\begin{aligned}
\alpha_1&=2\arcsin(x_a),
\\
\alpha_2&=2\arccos(\abs{X_4}),
\\
\alpha_3&=2\arccos[\abs{X_3} /\sin(\alpha_2/2)],
\\
\alpha_4&=2\arccos[\abs{X_2} /\sin(\alpha_2/2)/\sin(\alpha_3/2)].
    \label{Eq:prepare_angle}
\end{aligned} 
\end{equation} 

The experimental implementation of rotation direction change is achieved by controlling the phase of the microwave signal input to the acousto-optic modulator. If we define one phase as 0, then when a reversal of rotation is required, it is sufficient to set the phase to $\pi$. Since we are using a double-passed acousto-optic modulator, the actual phase for the microwave signal is either 0 or $\pi/2$. 

\section{\label{section2} Appendix C: Data readout}

\begin{figure*}[ht]
	\begin{center}

	\includegraphics[width=15.2cm]{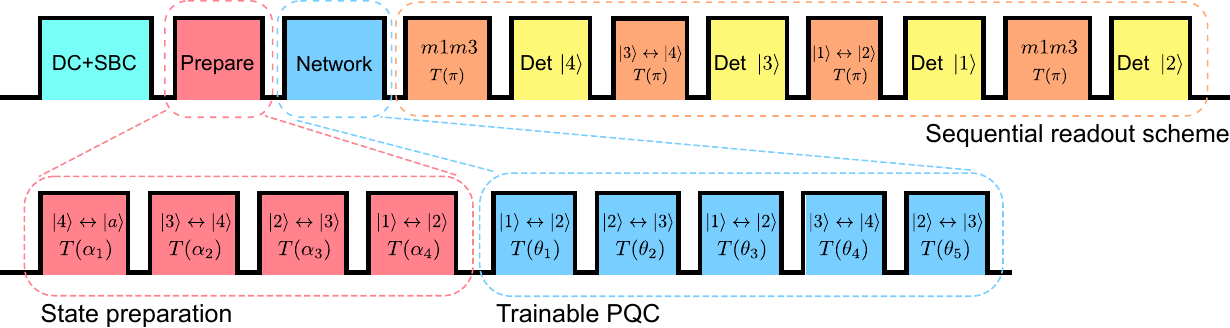}
	\caption{Pulse sequences for reading out populations of multi-levels of a single $^{40}\rm{Ca}^+$ ion. DC stands for Doppler cooling, SBC for sideband cooling, and Det for fluorescence detection. Especially, $m1m3$ stands for the transition of $\ket{2}\leftrightarrow \ket{D_{1/2}, -3/2}$. The preparation process consists of four 729~nm laser pulses carried out between four different pairs of energy levels. Note that, for error backpropagation during training, we need to measure the inner state $\ket{\boldsymbol{\zeta}^{k}}$, after applying different numbers of pulses. For the training process, we need to measure 6 times, but only once for the testing section of the experiment.
	}
	\label{fig:population_exp_sequence_1}
	\end{center}
\end{figure*} 

We now describe the process of reading out data from a $^{40}\rm{Ca}^+$ ion qudit. For clarity, we first define the representation of the 729~nm sub-transitions. Specifically, we combined the $m_j$ quantum numbers (without the $\frac{1}{2}$ coefficient) of the two levels to label the sub-transition, with the $4s~^2S_{1/2}$ state preceding the $3d~^2D_{5/2}$ state. For instance, the label $m1m3$ corresponds to the transition between the states $\ket{2}$ and $\ket{D_{5/2}, -3/2}$ ($m1$ means $-1$).

We employ a sequential readout scheme
to simultaneously read out the populations of all five qudit levels using a single pulse sequence. In a $^{40}\rm{Ca}^+$ ion, $4s~^2 S_{1/2}$ is the bright state, while $3d~^2D_{5/2}$ is the dark state. To measure the population of a certain energy level, it is necessary to transfer its population to a sub-level of the $4s~^2S_{1/2}$ manifold, and to shelve the population of unconcerned levels to the $3d~^2D_{5/2}$ manifold. In the sequential readout scheme, resonant 729~nm $\pi$ pulses are applied, and the populations are repeatedly redistributed across the system levels and ancillary levels, except for the level $\ket{D_{5/2},-5/2}$. This ensures that only one sub-level of the $4s~^2 S_{1/2}$ state is populated, followed by fluorescence detection.

Although four detections are carried out within a single experimental sequence, the results of each detection are not independent. The latter measurements depend on the results of the earlier ones. Specifically, the projection-valued measurements rely on the first brightness observed in the detections, while the subsequent measurements are conditioned on the previous detections being dark. The population measurement results, denoted as $\rho_i$, are described by the following relations, where $1$ represents a bright outcome, $0$ represents a dark outcome, and $*$ serves as a wildcard:
\begin{equation}
\begin{aligned}
&\rho_1=\frac{N(1***)}{N(****)},\rho_2=\frac{N(01**)}{N(****)},\\
&\rho_3=\frac{N(001*)}{N(****)},\rho_4=\frac{N(0001)}{N(****)},
\end{aligned}
    \label{Eq:detect_population_calculation}
\end{equation} 
where $N$ denotes the count that satisfies a given condition, and the total number of shots is $N(****) = 1500$. The subscript in $\rho_i$ indicates the detection number (not the qudit level label). The order in which populations are read out is $\ket{4}, \ket{3}, \ket{1}, \ket{2}$, as shown in Fig.~\ref{fig:population_exp_sequence_1}.

In the experiment, the readout procedure begins after cooling, state preparation, and the application of the trainable parameterized quantum circuit, as shown in Fig.~\ref{fig:population_exp_sequence_1}. The time allocated for Doppler cooling is $3~\rm{ms}$, while sideband cooling requires $0.7~\rm{ms}$, state preparation and network setup take $0.4~\rm{ms}$, and fluorescence detection lasts $0.3~\rm{ms}$. In a conventional approach, four independent experimental scripts would be required, each involving a cooling phase at the beginning. This leads to significant time overhead, as the cooling phase dominates the total sequence duration. By contrast, our sequential readout scheme consolidates the process into a single experimental script, thus reducing the time overhead significantly, even though the total experimental duration is extended. Moreover, the sequential scheme simplifies the experimental control, as only one script needs to be managed, rather than four.

Additionally, our experimental control system is optimized for the sequential readout scheme. We use the ARTIQ control system (provided by mlabs), which offers high flexibility in conditioning, multiple loops, and custom data storage formats.

\begin{figure}[ht]
	\includegraphics[width=9.0cm]{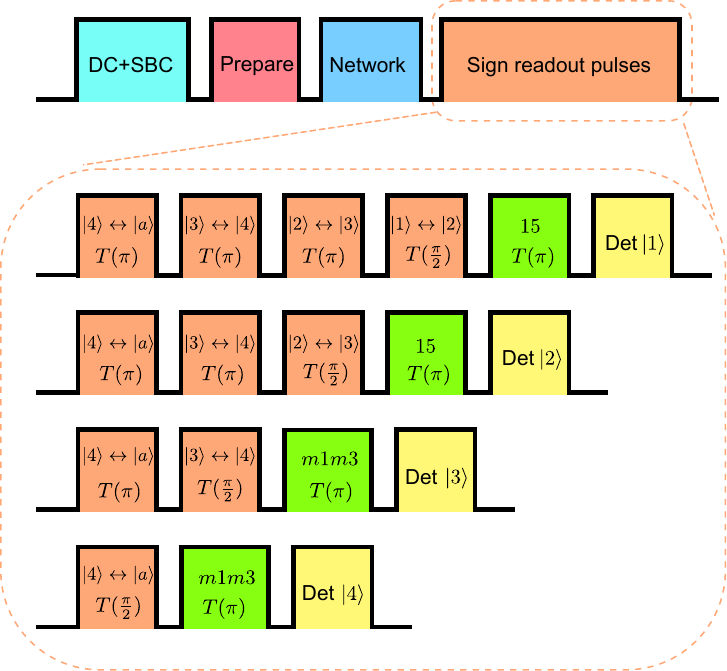}
	\caption{Pulse sequences for measuring the signs of the data. Cooling, preparing, and parameterized network pulses are the same as Fig.~\ref{fig:population_exp_sequence_1}. In which $15$ stands for the transition of $\ket{4}\leftrightarrow \ket{D_{1/2}, +5/2}$, and Det for fluorescence detection. $m1m3$ and $15$ are shelving transitions.
	}
	\label{fig:phase_exp_sequence_1}
\end{figure} 

The second part of the readout procedure involves determining the sign of each data element. This is achieved using four separate experimental scripts, as illustrated in Fig.~\ref{fig:phase_exp_sequence_1}. The sign determination procedure is inherently linked to the population readout results discussed earlier, aiming to reduce measurement overhead. Essentially, the sign of each qudit level is determined by measuring the interference between that level and the ancillary level $\ket{a}$.

Since the amplitudes of the data elements are unknown, interference must be induced between all four levels ($\ket{1}, \ket{2}, \ket{3}, \ket{4}$) and the ancillary level $\ket{a}$ using $\pi/2$ pulses. For the levels $\ket{1}, \ket{2}, \ket{3}$, which are not directly coupled, several $\pi$ pulses are needed as intermediaries to facilitate the interference, as shown in Fig.~\ref{fig:phase_exp_sequence_1}.

For brevity, we take $\ket{3}$ for example. Assume that we have already measured the population $\rho^{pop}_{3}$. Let $x_a^2 = 0.2$ denote the population of the ancillary level $\ket{a}$. The results of the sign readout can be expressed as:

\begin{equation}
\begin{aligned}
\text{if}\  \rho^{sign}_{3}=\frac{|\sqrt{\rho^{pop}_{3}}-x_a|^2}{2}, \text{then sign}(3)=-,
\\
\text{if}\  \rho^{sign}_{3}=\frac{|\sqrt{\rho^{pop}_{3}}+x_a|^2}{2},\text{then sign}(3)=+.
\label{Eq:Phase_two_results}
\end{aligned}
\end{equation}

These correspond to the ideal cases. In practice, we determine the sign by comparing the measured value $\rho^{sign}_{3}$ with the two ideal values. The sign is chosen based on the minimum distance between $\rho^{sign}_{3}$ and the two possible ideal outcomes. In the experiment, the sign readout is repeated $600$ times per level, resulting in a total of $2400$ repetitions for all four levels, as compared to the $1500$ repetitions required for population readout.

\section{\label{section3}Appendix D: Error caused by shot noise}%

During a given fluorescence detection period, when the ions are in the $S_{1/2}$ or $D_{5/2}$ manifold, the photon counts detected by the photomultiplier tube follow a Poisson distribution with respective mean values of $\lambda=36.4$ or $2.7$ for the 
$S_{1/2}$ and $D_{5/2}$ states. We define a detection threshold $N_{th}=15$, where the ion is considered to be in the $S_{1/2}$ manifold if the average photon count exceeds $N_{th}$, and in the $D_{5/2}$ manifold if it falls below $N_{th}$.

The population readout result can be obtained by performing $N$ independent measurements, as given by $\rho=\frac{\sum^N_i K_i}{N}=\frac{Z}{N}$, where $K_i$ are independent, identically distributed random variables following a binomial distribution with parameter $p$, and $Z$ is the total sum of these random variables. We can then compute the mean and variance of the population measurement as $E(\rho)=\frac{E(Z)}{N}=p$ and $D(\rho)=\frac{E(Z^2)-(E(Z))^2}{N^2}=\frac{p(1-p)}{N}$. The standard deviation of the population readout results is therefore $\sigma(\rho)=\sqrt{\frac{p(1-p)}{N}}$.

Next, we examine the standard deviation of the absolute value of the amplitude of the data vector. Let the amplitude be represented as $x=\sqrt{\rho}$. Using the uncertainty propagation formula, the standard deviation of $x$ is derived as:
\begin{equation}
\sigma(x)=\frac{\frac{1}{2}\sqrt{\rho}\sigma(\rho)}{\rho}=\frac{\sqrt{1-p}}{2\sqrt{N}}.
\label{Eq:Population_ReadoutError_3}
\end{equation}

In practice, we take 1500 shots to measure the populations, which results in an estimated maximum standard deviation for the amplitude magnitude of $max(\sigma(x))=\frac{1}{2\sqrt{N}}=0.013$.

\section{\label{section4}Appendix E: Residual analysis}
\begin{figure}[ht]
\centering
    \includegraphics[width=0.99\columnwidth]{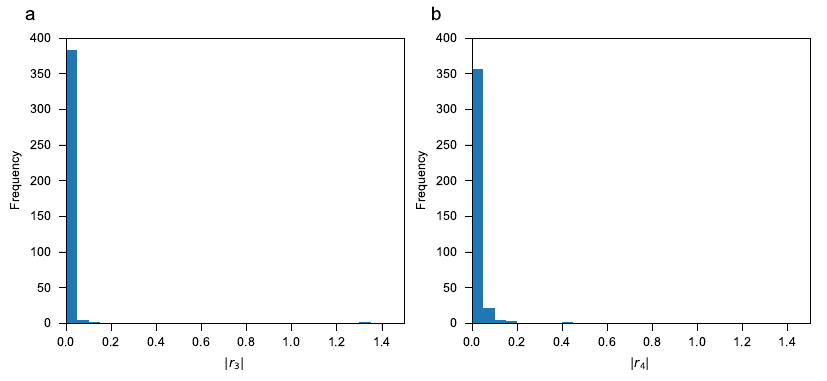}
  \caption{Residual distributions of (a) $y_3$ and (b) $y_4$. The $r_3$ and $r_4$ are defined by $r_i = y_i-\hat{y_i}~(i=3,4)$, in which $\hat{y_i}$ denotes the theoretical value.}
  \label{fig:residual_analyze}
\end{figure}

Here we provide a residual analysis of the amplitudes $y_3$ and $y_4$, which are used for sample classification. For the $j$th test sample, the residual is defined as $r_i^{(j)}=y_i^{(j)}-\hat{y}_i^{(j)}$ ($i=3,4$), where $y_i^{(j)}$ and $\hat{y}_i^{(j)}$ denote the experimental and theoretical values, respectively. For the test set comprising 395 samples, the distributions of residuals $r_3$ and $r_4$ are presented in Fig.~\ref{fig:residual_analyze}. The residuals of most samples are concentrated near zero, indicating overall agreement between experiment and theory.
The root mean squared error (RMSE) of $y_3$ and $y_4$ are defined by $RMSE(y_i)=\sqrt{\frac{1}{n}\sum^{n}_{j=1}|r_i^{(j)}|^2}, i=3,4$. Here, $n=395$ denotes the number of samples in the test set. The resulting values of $RMSE(y_3)$ and $RMSE(y_4)$ are $1.32\times 10^{-1}$ and $9.11\times 10^{-2}$, respectively.

\section{\label{section5}Appendix F: Simulations of multiple noises}

\begin{figure}[ht]
\centering
    \includegraphics[width=0.99\columnwidth]{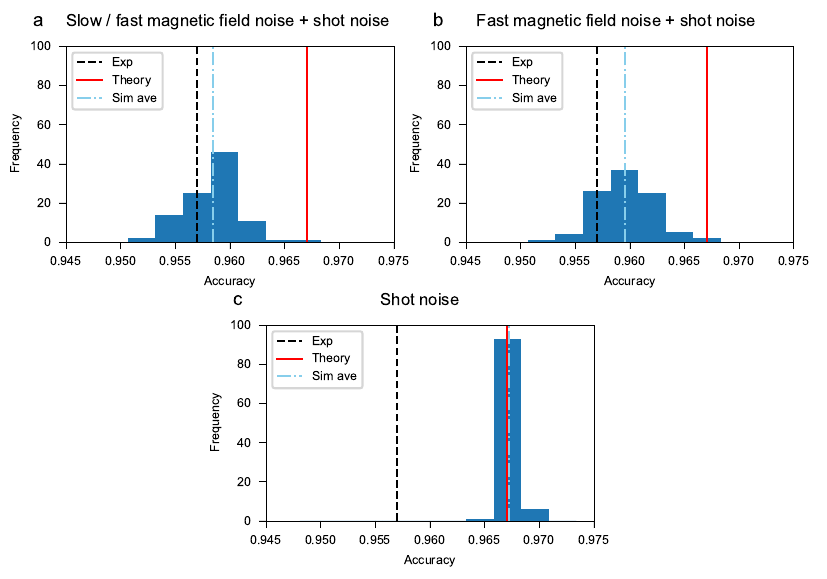}
  \caption{Simulations of the effects of slow and fast magnetic field noise and shot noise on test-set classification accuracy. Histograms are generated from 100 independent simulation runs under identical conditions for each noise configuration. (a) All three noise sources included. (b) Fast magnetic field noise and shot noise included. (c) Shot noise only. The red solid lines indicate the theoretical accuracy of 96.7\%, while the black dashed lines mark the experimental accuracy of 95.7\%. The light blue dash-dotted lines represent the mean accuracy averaged over the 100 simulation runs.}
  \label{fig:magnetic_analyze}
\end{figure}

In this section, we provide a detailed simulation considering multiple types of noise. Slowly and rapidly fluctuating environmental magnetic field noises affect the experiment in different manners. In our setup, the rapidly fluctuating magnetic field noise is the dominant factor causing spin decoherence of the ion. Meanwhile, owing to large electrical equipment in the vicinity of the experimental setup, the magnetic field along the ion quantization axis undergoes slow variations of up to 0.36~mGs, as verified by calibrations of the 729~nm transition resonance.

We simulate the single-qudit QNN using the Lindblad Master equation, taking into account three distinct types of noise: slow magnetic field fluctuations, spin decoherence, and shot noise.
We employ a sampling method to simulate shot noise, with the number of sampling instances consistent with the experimental values of 1500 and 600. We approximate the slow magnetic field fluctuation as a two-point distribution of +0.18~mGs and -0.18~mGs. In addition, the qubit coherence time $T_2$ is approximately 4 ms (for the $\ket{1}\longleftrightarrow\ket{2}$ and $\ket{3}\longleftrightarrow\ket{4}$ transitions, with different values for other transitions), based on which we configure the Lindblad operators.

The simulation results are shown in Fig.~\ref{fig:magnetic_analyze}, consisting of 100 independent runs. Simulation results show that, in the presence of shot noise alone (Fig.~\ref{fig:magnetic_analyze}c), the classification accuracy on the test set is close to the theoretical value. When both shot noise and spin decoherence (fast magnetic field noise) are considered (Fig.~\ref{fig:magnetic_analyze}b), the test-set accuracy decreases to 95.9\%. Incorporating all three types of noise leads to a further slight reduction in the test-set accuracy (Fig.~\ref{fig:magnetic_analyze}a). Given that the experimentally measured test-set accuracy is 95.7\%, the discrepancy between the experimental results and the theoretical values can be primarily attributed to spin decoherence induced by rapidly fluctuating magnetic field noise.

\section{\label{section6}Appendix G: Details of baseline comparisons}

\begin{figure}[ht]
\centering
    \includegraphics[width=0.99\columnwidth]{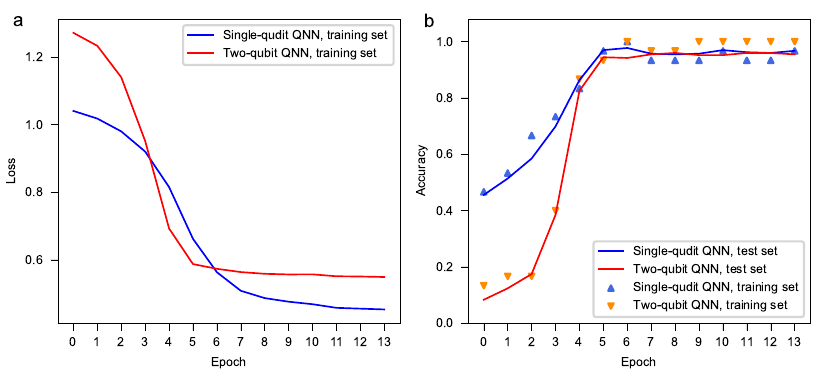}
 \caption{(a) Average training loss. The red line represents the two-qubit QNN, and the blue line represents the single-qudit QNN. The loss at each epoch is averaged over all 30 training samples, while the gradients are evaluated using the 15 samples randomly reselected at each epoch. (b) Classification accuracy. Dots denote the training-set accuracy, while solid lines denote the test-set accuracy.}
  \label{fig:baseline_comparison}
\end{figure}

\begin{figure}[ht]
\centering
    \includegraphics[width=0.60\columnwidth]{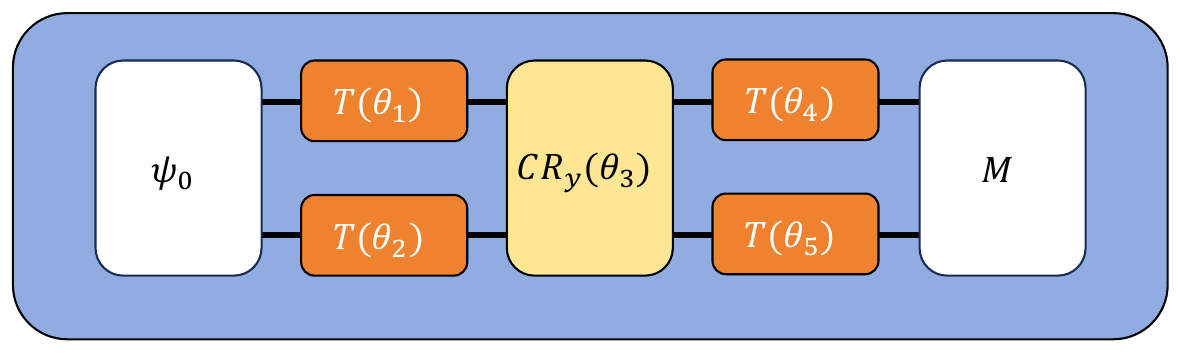}
  \caption{Schematic circuit diagram of the two-qubit QNN. The definition of the $T(\theta)$ gate is the same as that in the main text, and $CR_y$ denotes the controlled-$R_y$ gate. The four-dimensional PCA feature vector is amplitude-encoded as $\ket{\psi_0}=x_1\ket{00}+x_2\ket{01}+x_3\ket{10}+x_4\ket{11}$. This state then evolves through five trainable gates, after which the amplitudes of the four computational-basis states are read out. The classification decision is based on the amplitudes of the $\ket{00}$ and $\ket{11}$ states.}
  \label{fig:two_qubit_qnn}
\end{figure}

In this section, we provide details about the baseline comparisons. All models were trained on the same 30 training samples and subsequently evaluated on the same 395 test samples. Each sample was represented by the same four-dimensional PCA features used in the experiment. The trained linear support vector machine achieves a training accuracy of 100\% and a test accuracy of 97.2\%. The noise-free numerical simulations of the two-qubit QNN and a single-qudit QNN are shown in Fig.~\ref{fig:baseline_comparison}. Both QNNs were trained independently in the numerical simulations. The two-qubit QNN consists of four $T(\theta)$ gates and one controlled $R_y(\theta)$ gate, as shown in Fig.~\ref{fig:two_qubit_qnn}. The four-dimensional PCA feature vector is amplitude-encoded to the two-qubit input state $\ket{\psi_0}=x_1\ket{00}+x_2\ket{01}+x_3\ket{10}+x_4\ket{11}$, where $x_i$ are the elements of the 4-dimensional data vector after PCA. During training, we used the same mini-batch strategy by randomly selecting 15 of the 30 training samples to compute the gradient at each epoch. The two-qubit QNN was trained using the parameter-shift rule. After 13 training epochs, the two-qubit QNN achieves a training set accuracy of 100\% and a test accuracy of 95.4\%, while the noise-free single-qudit QNN achieves a training set accuracy of 96.7\% and a test accuracy of 96.7\%. Overall, the three model classes exhibit comparable test performance on this small-scale proof-of-principle task, with their accuracies differing by less than two percentage points.

\end{document}